\documentclass[aps,prl,preprint,groupedaddress]{revtex4}
\usepackage{graphicx}

\begin{document}


\title{The Statistics of DNA Capture by a Solid-State Nanopore}


\author{Mirna Mihovilovic}
\author{Nick Hagerty}
\author{Derek Stein}
\email[]{derek_stein@brown.edu}
\affiliation{Physics Department, Brown University, Providence, RI, USA}


\date{\today}

\begin{abstract}
A solid-state nanopore can electrophoretically capture a DNA molecule and pull it through in a folded configuration.  The resulting ionic current signal indicates where along its length the DNA was captured.  A statistical study using an $8$\,nm wide nanopore reveals a strong bias favoring the capture of molecules near their ends. A theoretical model shows that bias to be a consequence of configurational entropy, rather than a search by the polymer for an energetically favorable configuration.  We also quantified the fluctuations and length-dependence of the speed of simultaneously translocating polymer segments from our study of folded DNA configurations.
\end{abstract}


\maketitle



A voltage-biased nanopore is a single-molecule detector that registers the disruption of $I$, the ionic current through the nanopore, caused by the insertion of a linear polyelectrolyte \cite{KBBD, Li, Dekker}. Most previous studies have focused on instances where the nanopore electrophoretically captures DNA at one end and then slides it through in a linear, head-to-tail fashion.  However, a $\approx10$ nm-wide solid-state nanopore can also capture DNA some distance from its end and pull it through in a folded configuration \cite{Naturemat, Chen2004NanoLett, Storm2005PRE}.  Folded DNA translocations entail the simultaneous motion of multiple segments through the nanopore, which may exhibit cooperative behavior that alters the translocation dynamics \cite{Melchionna2009PRE}. The mechanical bending energy associated with folds may influence the capture of DNA \cite{Muthu2007JCP}.  Importantly, the study of folded configurations provides snapshots of molecules at the moment of insertion, which offer clues about how the nanopore captures them from solution. The capture process is relevant to applications of nanopores that seek to extract sequence-related information from unfolded molecules.

\begin{figure}[htb]
\begin{center}
\includegraphics[scale=0.3]{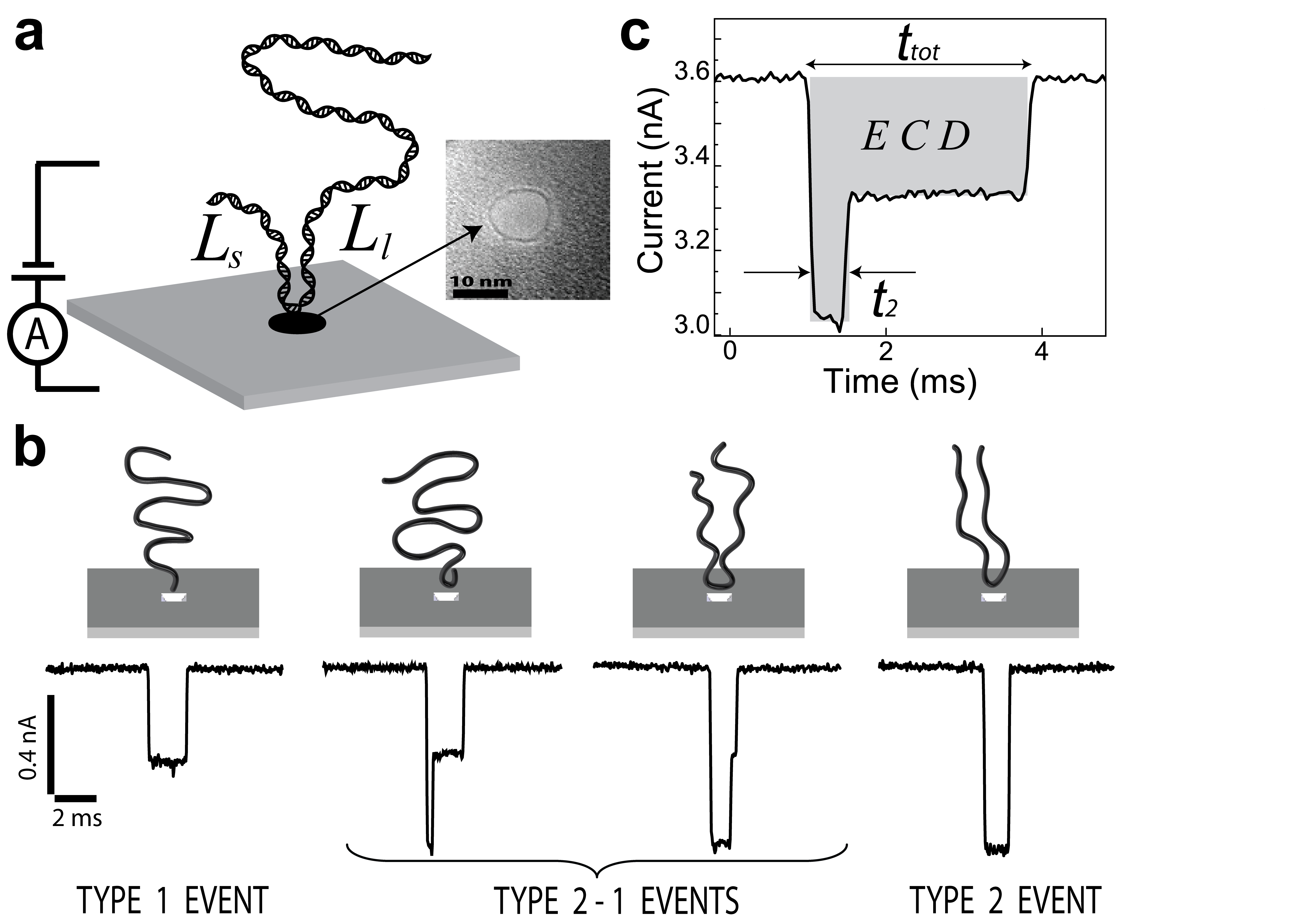}
\end{center}
\caption{a) A nanopore captures DNA from solution and initiates electrophoretic translocation by forming a hairpin. Segments of length $L_l$ and $L_s$ extend from the capture location. (Detail) TEM image of the 8\,nm wide nanopore used. b) Ionic current traces from translocation events of type 1, 2-1, and 2 indicate the capture location. c)  The ionic current trace of a folded DNA molecule shows $t_2$, $t_{\text{tot}}$, and ECD.}
\label{fig1}
\end{figure}

When DNA encounters a nanopore, the electrophoretic force can initiate translocation by inducing a hairpin fold in the molecule that protrudes into the nanopore.  Two segments of DNA extend from the initial fold, a long one of length $L_l$ and a short one of length $L_s$ (Fig.\,\ref{fig1}(a)). The capture location, $x\equiv \frac{L_s}{L_s+L_l}$, is the fractional contour distance from the initial fold to the nearest end. The time for each segment to translocate is measurable from the time trace of $I$ \cite{Naturemat, Chen2004NanoLett,Storm2005PRE} and can be used to estimate $x$. Storm \emph{et al.} inferred the distribution of $x$ for $\lambda$\,DNA translocations and concluded that folds occur with equal probability everywhere along a molecule's length, but that the DNA is more likely to be captured at its ends because of the lower energetic cost of threading an unfolded molecule \cite{Storm2005PRE}. This implies that molecules test multiple configurations prior to capture, which is a statistical process governed by energetic considerations.  By contrast, Chen \emph{et al.} reported a bias for unfolded translocations that increased with applied voltage \cite{Chen2004NanoLett}. This finding implies that molecules pre-align in the fields outside the pore rather than sample multiple configurations prior to capture.  No model for the distribution of $x$ is available to help evaluate these competing pictures.

Here, we present a study of DNA translocations of an 8\,nm-wide solid-state nanopore which reveals a strongly biased distribution of capture locations, where the probability of capture increases continuously and rapidly towards the DNA's ends. The equilibrium distribution of polymer configurations outside the nanopore offers a natural explanation for this surprising finding. We present a simple but successful model of that distribution in which only the configurational entropy is important. Finally, we show that a constant mean translocation velocity and Gaussian velocity fluctuations explain the translocation dynamics of folded DNA well, but that a weak length-dependence of the mean segment velocity exists.

The 8\,nm diameter solid-state nanopore we used (Fig.\,\ref{fig1}(a), detail) was fabricated in a 20\,nm-thin low-stress silicon nitride membrane following procedures described elsewhere \cite{fabrication}. The nanopore bridged two fluid reservoirs containing degassed aqueous 1 M KCl, 10 mM Tris-HCl, 1 mM EDTA buffer (pH 7.7). An electrometer (Axon Axopatch) applied 100\,mV across the nanopore and monitored $I$ using two Ag/AgCl electrodes immersed in the reservoirs. A $10$\,kHz, 8-pole, low-pass Bessel filter conditioned $I$ prior to digitization at 50\,kilo-samples per second.  The open-pore current was $I=3.6$\,nA. After adding $\lambda$ DNA (16.5\,$\mu$m long, New England Biolabs) to the negatively charged reservoir at a concentration of 24\,$\mu$g/mL, transient blockages in $I$ were observed, such as the ones shown in Fig.\,\ref{fig1}(b).

\begin{figure}[htb]
\begin{center}
\includegraphics[scale=1]{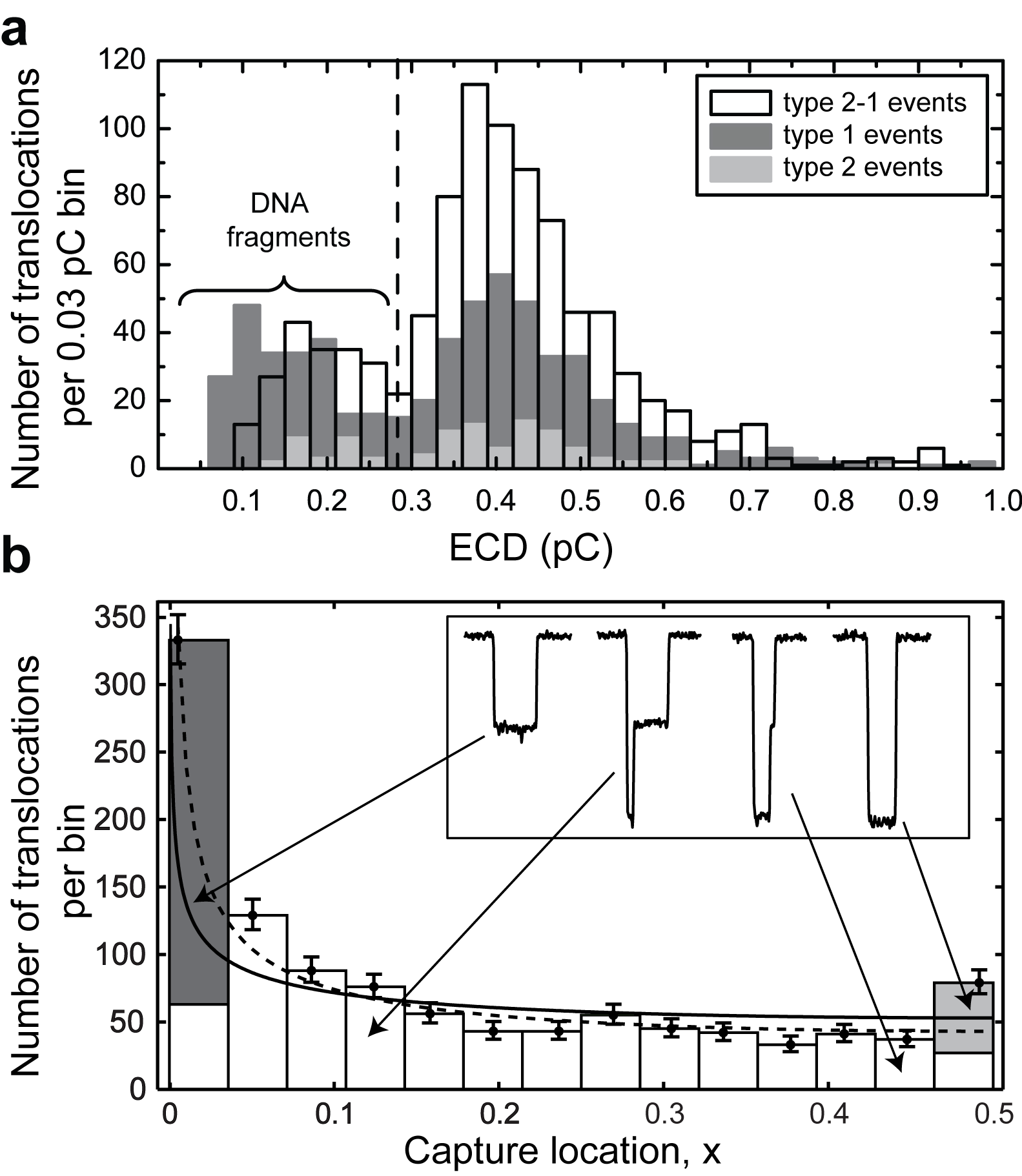}
\end{center}
\caption{a) Overlaid ECD distributions for translocations of type 1 (dark grey), 2-1 (white), and 2 (medium grey). Events with ECD $< 0.27$ pC and six with ECD $>3$ pC were dropped from subsequent analyses in order to exclude fragmented and stuck DNA molecules, respectively. Only $0\text{\,pC}\leq\text{ECD}\leq1$\,pC is plotted for clarity. b) Distribution of capture locations.  The stacked histogram bars indicate the number of events of each type in a bin. Data points indicate the total number of events of all types and their mean $x$ in a bin. Error bars indicate the square root of the total events. The distributions predicted by Eq.\,\ref{probability} are shown for the theoretical $\gamma=0.70$ (solid line) and for the weighted best fit $\gamma=0.46$ (dashed line).
}
\label{fig2}
\end{figure}

The blockages show quantized steps in $I$ that indicate where the nanopore captured each molecule, as illustrated in Fig.\,\ref{fig1}(b). Unfolded molecules decreased $I$ by $\approx0.278$\,nA for the full duration of the translocation event, $t_{\text{tot}}$.  We call these ``type 1'' events.  Folded molecules cause two segments to occupy the nanopore simultaneously, thereby doubling the reduction in $I$ for a time $t_2$. Two segments occupied the nanopore for the full duration of ``type 2'' events, indicating molecules captured at the midpoint. A transition from double to single occupancy was observed in ``type 2-1'' events, indicating molecules captured somewhere between an end and the midpoint. Fig.\,\ref{fig2}(c) shows a type 2-1 event that illustrates $t_{\text{tot}}$ and $t_2$; we judged the occupancy of the nanopore to have changed when $I$ rose or fell 80\,\% of the way to the next blockage level. We also observed event types which indicate molecules captured and folded by the nanopore at multiple locations. For the present study, however, we restrict our attention to translocations with at most a single fold, which account for $\sim 70 \%$ of all events.

We found evidence that a minority of the current blockages were caused by fragments of $\lambda$ DNA that we wish to exclude from further analysis.  We considered the event charge deficit (ECD), which is the current blockage integrated over the duration of an event (illustrated in Fig.\,\ref{fig1}(c)). Fig.\,\ref{fig2}(a) plots the ECD distributions for events of type 1, 2-1 and 2.  Most events fall into the main peaks that are centered at $0.408\pm0.003$\,pC, regardless of the event type. We attribute those events to intact $\lambda$-DNA molecules \cite{Naturemat}. Minor peaks in the distributions near $0.15$ pC likely correspond to fragments of those molecules.  To obtain a monodisperse ensemble, we excluded all events with ECD\,$< 0.27$\,pC from further analysis.  We also excluded six events with ECD$>3$\,pC, presumably caused by molecules that stuck to the nanopore.  These restrictions leave us with an ensemble of $\sim 1100$ identical $\lambda$ DNA molecules that translocated with at most a single fold. 

For each translocation event, we obtained the capture location, $x$, by assuming that the translocation speed, $v$, was constant over the duration of the event, which follows the approach of Storm \emph{et al.}  \cite{Storm2005PRE} and gives:
\begin{equation}
x= \frac{t_2}{t_2+t_{\text{tot}}}.
\label{define x}
\end{equation}
Below we shall investigate the accuracy of that assumption and explore the consequences of fluctuations and a contour length dependence in $v$.

Figure\,\ref{fig2}(b) presents a histogram of the capture locations. We selected a bin size that avoids a possible artifact of the limited measurement bandwidth; since there is a lower bound on $t_2$, it would be difficult to populate bins near $x=0$ if the bin size were too small.  The distribution shows that the frequency of capture was highest near $x=0$, decreasing rapidly but smoothly with distance away from the ends, and becoming a slowly decreasing function of $x$ near $x=0.5$.  The bin that includes $x=0.5$ rises above the trend. 

We propose a physical model to explain the distribution of capture locations. We assume that a DNA molecule has enough time to sample all available configurations as it approaches the nanopore. At the moment of capture, the nanopore randomly selects a configuration from the equilibrium ensemble. We model that configuration as a pair of independent self-avoiding walks (SAWs) of lengths $L_s$ and $L_l$, tethered to the surface at a single point representing the nanopore. We discuss these assumptions below.

For a single polymer, the total number SAWs of length $L$, $\Omega(L)$, has the following asymptotic form \cite{DeGennesScaling}:

\begin{equation}
\Omega(L)\sim \mu^L L^{\gamma-1}.
\label{microstates}
\end{equation}

\noindent  $\gamma$ is a universal scaling exponent which depends solely on the dimensionality of the lattice and $\mu$ is the lattice coordination number. Barber \emph{et al.} studied SAWs tethered to a surface and obtained $\gamma \approx 0.70$ from simulations on a cubic lattice \cite{Guttman}.

The number of configurations available to a molecule captured at $x$, $\Omega_{L}(x)$, is the product of the number of SAWs for each segment, $\Omega(L_s)$ and $\Omega(L_l)$. From $L_s + L_l = L$ and Eq.\,\ref{microstates}, it follows that $\Omega_{L}(x) = \Omega(L_s) \cdot \Omega(L-L_s)$. The probability of capturing a molecule at $x$, $P(x)$, is proportional to $\Omega_L(x)$, therefore we find:
\begin{equation}
P(x)= Ax^{\gamma-1}\cdot(1-x)^{\gamma-1}.
\label{probability}
\end{equation}
The solid line in Fig.\,\ref{fig2}(b) plots the distribution of capture locations predicted by Eq.\,\ref{probability} for $\gamma=0.7$. The proportionality constant $A$ was obtained from a weighted least squares fit to the data.  By contrast, the best fit of Eq.\,\ref{probability} when $\gamma$ is left as a free parameter, indicated by the dashed line in Fig.\,\ref{fig2}(b), obtains $\gamma=0.46\pm0.03$.

The two-tethered-polymer model describes the observed distribution of capture locations well. Note that the skewness arises naturally from configurational entropy alone; every DNA configuration is represented with equal probability and there is no need to invoke a bending energy, as Storm \emph{et al.} did, to explain the preponderance of molecules captured near their ends \cite{Storm2005PRE}. The model disagrees most significantly with the data at $x=0.5$, where more events were observed than predicted. That discrepancy can be explained by the translocation of circular $\lambda$\,DNA molecules, whose complementary single-stranded ends had bound, resulting in extra type 2 events.  An important implication of our model is that DNA does not search for an energetically favorable configuration before initiating a translocation.

A question that our experiments cannot address is where, in relation to the nanopore, the capture location is determined. Within our model, $x$ is determined at the nanopore; however, recent studies have identified a critical radius from the nanopore, typically on the scale of hundreds of nanometers, within which electrophoretic forces overwhelm diffusion \cite{Gershow, Grosberg}. It is possible that the first segment to insert is transported essentially deterministically to the nanopore from some distance away without altering the distribution of $x$. Similarly, our assumption that a DNA molecule is at equilibrium prior to capture is not seriously compromised if the molecule becomes stretched out of equilibrium by the field gradients only after the capture location has been determined. The forces on DNA beyond the nanopore may restrict the available configurations and thereby reduce $\gamma$.

A third assumption of our model worth considering is that both segments of the captured polymer behave independently. In addition to undergoing self-avoiding walks, both segments should avoid one another. Theoretically, $\gamma$ decreases to $\approx 0.60$ when two segments of equal length are tethered to the same point on a surface \cite{starfish}.

\begin{figure}[htb]
\begin{center}
\includegraphics[scale=1.1]{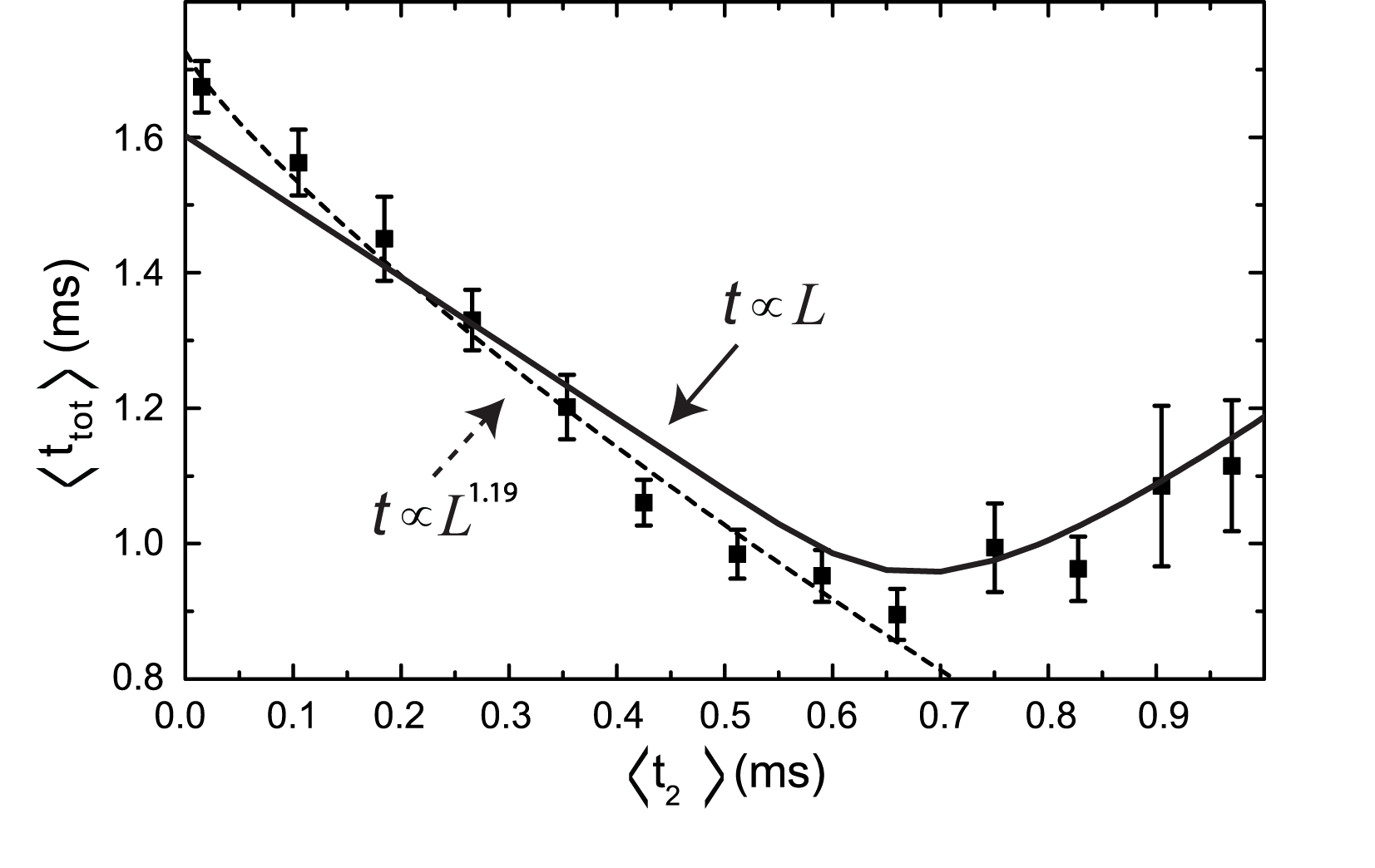}
\end{center}
\caption{Dependence of $\langle t_{\text{tot}}\rangle$ on $\langle t_2\rangle$. Error bars indicate the standard deviation of the mean in a 80\,$\mu$s bin. Bins with $\langle t_2\rangle>1$\,ms contain an insignificant number of events ($\leq2$).  The solid line shows the predictions of the dynamical model that includes velocity fluctuations described in text. The dashed line accounts for the length-dependence of the translocation speed of each segment with $t\propto L^{\alpha}$. The scaling exponent $\alpha=1.19\pm0.04$ was obtained from a weighted least squares fit to the data in the range $\langle t_2\rangle<0.7$\,ms.}
\label{fig3}
\end{figure}

We next turn to the translocation dynamics of folded molecules.  We estimated $x$ for each event by assuming that both segments translocated at the same speed; however, that assumption ignores fluctuations in the speed and any dependence on the length of a segment, which are both established features of unfolded DNA translocations \cite{Storm2005NanoLett,Golovchenko2011BJ}. In order to investigate our assumption in more detail, we divided the translocation data into 80\,$\mu$s bins of $t_2$. For each bin, $\langle t_{\text{tot}}\rangle$ and its standard deviation were calculated and plotted against $\langle t_2\rangle$ (Fig.\,\ref{fig3}). $\langle Q \rangle$ denotes the mean of quantity $Q$ in a 80\,$\mu$s bin. If both segments translocated at the same speed, we would expect $\langle t_{\text{tot}}\rangle$ to decrease in proportion with any increase in $\langle t_2\rangle$.  Fig.\,\ref{fig3} shows that $\langle t_{\text{tot}}\rangle$ in fact decreased approximately linearly with $\langle t_2 \rangle$ until $\langle t_2 \rangle \approx 0.7$\,ms, where $\langle t_{\text{tot}}\rangle$ began to rise. That turning point coincides approximately with the mean translocation time for type 2 events. 

The upswing in $\langle t_{\text{tot}}\rangle$ with $\langle t_2\rangle$ is the result of fluctuations in the translocation speed, as the following dynamical model illustrates. Consider a folded molecule whose two segments translocate with the same Gaussian distribution of speeds, $G_{v_0,\Delta v}(v)$. $v_0$ is mean translocation speed and $\Delta v$ is the standard deviation, which accounts for fluctuations.  Accordingly, if a segment translocates in a time $t_2$, the probability that its length was between $L_s$ and $L_s +dL_s$ is given by:
\begin{equation}
P\left(L_s \mid t_2 \right)dL_s \propto G_{v_0,\Delta v_0}\left(\frac{L_{s}}{t_{2}}\right) \frac{dL_{s}}{t_{2}}. 
\label{PofLsgivenT2}
\end{equation}
The probability distribution $P\left(L_s \mid t_2 \right)dL_s$  is normalized by integrating over $L_s$ from $0$ to $L$. The complementary segment has length $L_l=L-L_s$.  The probability that it takes between $t_{\text{tot}}$ and $t_{\text{tot}}+dt_{\text{tot}}$ to translocate is:
\begin{equation}
P\left(t_{\text{tot}} \mid L_s \right)dt_{\text{tot}} \propto G_{v_0,\Delta v_0}\left(\frac{L-L_{s}}{t_{\text{tot}}}\right)\frac{L-L_s}{t_{\text{tot}}^2} dt_{\text{tot}}.
\label{PofTtotgivenLs}
\end{equation}
Combining Eqs.\,\ref{PofLsgivenT2} and \ref{PofTtotgivenLs}, we find that when one segment translocates in a time $t_2$, the complementary segment will translocate in a time between $t_{\text{tot}}$ and $t_{\text{tot}}+dt_{\text{tot}}$ with a probability given by:
\begin{equation}
P\left(t_{\text{tot}} \mid t_2 \right)dt_{\text{tot}} \propto \left(\int_0 ^{L} P\left(t_{\text{tot}} \mid L_s \right)P\left(L_s \mid t_2 \right)dL_s\right)dt_{\text{tot}}.
\label{distribution}
\end{equation}
The distribution $P\left(t_{\text{tot}} \mid t_2 \right)$ is normalized by integrating over $t_{\text{tot}}$ from $t_2$ to $\infty$. A least squares fit of Eq.\,\ref{distribution} to the data in the first bin of Fig.\,\ref{fig3} ($\langle t_2 \rangle=0.015$) obtains $v_0 = 10.76 \pm 0.06$\,mm/s and $\Delta v/v_0 = 0.198 \pm 0.005$. With those parameters and Eq.\,\ref{distribution}, we calculated $\langle t_{\text{tot}} \rangle$ as a function of $t_2$ and plotted the results in Fig.\,\ref{fig3}. The predicted relationship agrees well with the data. 

Importantly, the dynamical model demonstrates the robustness of our method for obtaining the distribution of $x$ in Fig.\,\ref{fig2}(b).  Fluctuations lead to errors in estimating $x$ for a particular event, as one segment may translocate faster or slower than the other; however, the relationship between $t_{\text{tot}}$ and $t_2$ is the same on average as if there were no fluctuations.  Events with $t_2 >0.7$\,ms are drawn from tails of the speed distributions; $\langle t_{\text{tot}} \rangle$ rises with $\langle t_2 \rangle$ because both segments of molecules captured at $x \approx 0.5$ translocated more slowly than average, not because the segments translocated at different speeds on average. Accordingly, we found $x\approx 0.5$ for those events.

Finally, the slope of the data in Fig.\,\ref{fig3} for $\langle t_2\rangle <0.7$\,ms reveals a weak dependence of the translocation speed on the length of a segment. Long molecules are known to translocate more slowly than short ones in unfolded configurations \cite{Storm2005NanoLett} because the moving segment is longer and experiences more viscous drag when it is drawn to the nanopore from a large coil \cite{Grosberg2006,Golovchenko2011BJ}. Storm \emph{et al.} assumed a power law relationship between the translocation time and the length of unfolded DNA, $t\sim L^{\alpha}$, and found that the scaling exponent $\alpha=1.27$ \cite{Storm2005PRE}.  Assuming that each segment of a folded molecule obeys a similar scaling relationship and using $L_s + L_l = L$, we find that $t_{\text{tot}}=(t_{1}^{1/\alpha} - t_{2}^{1/\alpha})^{\alpha}$, where $t_{1}$ is the translocation time of unfolded molecules.  We fitted that expression to the data in Fig.\,\ref{fig3} for $\langle t_2\rangle <0.7$\,ms to obtain $\alpha=1.19\pm0.04$. Accounting for the length-dependent speed in estimating $x$ skews the distribution, raising the best fit exponent to $\gamma = 0.72 \pm 0.02$, which is closer to the theoretical value.

In conclusion, we measured the distribution of capture locations along $\lambda$ DNA molecules by an 8\,nm wide solid-state nanopore and presented a theoretical model which explains that distribution.  Surprisingly, the strong bias for capturing molecules near their ends is a consequence of the configurational entropy of the approaching polymer; molecules do not search for an energetically favorable configuration before translocating. We also used folded DNA configurations to probe the dynamics of multiple polymer segments translocating a nanopore simultaneously, thereby quantifying the fluctuations and the length dependence of the translocation speed.

\begin{acknowledgments}
The authors thank Z. Jiang, S.-C. Ying, X.S. Ling, L. Theogarajan, O. Elibol, and J. Daniels for useful discussions.  This work was supported by Intel Corporation and the National Science Foundation under Grant Number CBET-0846505.
\end{acknowledgments}

\bibliography{references}

\end{document}